\documentclass[a4paper,10pt]{article}
\usepackage[latin1]{inputenc}
\usepackage{graphics}
\usepackage{url}
\usepackage{cite}
\usepackage{pdfpages}
\begin{document}
\title{CIDS country rankings: comparing documents and citations of USA, UK and China top researchers}
\author{
Francisco M. Couto
\\
\mbox{ }
\\ \small
LaSIGE, University of Lisbon, Portugal
}
\date{}
\maketitle
\begin{abstract}
This technical report presents a bibliometric analysis of the top 30 cited researchers from USA, UK and China.
The analysis is based on Google Scholar data using CIDS. 
The researchers were identified using their email suffix: edu, uk and cn.
This na\"{i}ve approach was able to produce rankings consistent with the SCImago country rankings using mininal resources in a fully automated way. 
\end{abstract}
\section{Introduction}
\label{intro}
The research impact and relevance of scientists, institutions and even countries
are being measured and analysed by multiple bibliometric studies~\cite{lopez2009comparing, jacso2010comparison, aguillo2012google}.

SCImago country rank reports the number of citations, self-citations,
citations per document and h-index for the period of 1996-2007~\cite{scimago2007sjr}.
The top three countries are USA (United States of America), China and UK (United Kingdom).  
USA has the highest number of documents and citations, China the second highest 
number of documents and UK the second highest number of citations. 

This report presents a bibliometric analysis that was performed to have an idea of these numbers for USA, UK, and China
but based on Google Scholar\footnote{\url{http://scholar.google.com}} data by using 
CIDS (Citation Impact Discerning Self-citations), a tool
that automates the post-processing of raw Google Scholar data\cite{couto2009handling}.

\section{Methodology}
The first step was to obtain the list of top researchers of each country. 
This was done by searching the Google Scholar profiles by the following keywords: 
edu\footnote{\url{http://scholar.google.pt/citations?view_op=search_authors&mauthors=.edu}} for USA, 
uk\footnote{\url{http://scholar.google.pt/citations?view_op=search_authors&mauthors=.uk}} for UK, 
and cn\footnote{\url{http://scholar.google.pt/citations?view_op=search_authors&mauthors=.cn}} for China. 
From the results, for each country 
the first 30 profiles which email ended with the suffix edu, uk, or cn, respectively, were selected. 
This is a rough estimate since not all the researchers from these three countries have an email
with any of those suffixes. However, it is expected to give us the big picture.

The list of profiles was given as input to CIDS, a freely available tool that automatically 
calculates bibliometric parameters based on Google Scholar for teams of researchers~\cite{fcouto:tr2013}.

\section{Results}

\begin{table}[!h]
\centering
\resizebox{0.95\textwidth}{!}{ \begin{tabular}{l|rrrrr}
Country & Citable documents & Citations & Self Citations & Cits per Doc & H index \\
\hline \hline
USA  & 6,672,307 & 129,540,193 & 62,480,425 & 20.45 & 1,380 \\
China  & 2,655,272 & 11,253,119 & 6,127,507 & 6.17 & 385 \\
UK  & 1,763,766 & 31,393,290 & 7,513,112 & 18.29 & 851 \\
\end{tabular} }
\caption{SCImago country rankings.}
\label{tab:scimago}
\end{table}

\begin{table}[!h]
\centering
\resizebox{0.95\textwidth}{!}{ \begin{tabular}{l|rrrrr}
Country & Citable documents & Citations & Self Citations & Cits per Doc & H index \\
\hline \hline
USA  & 6,877 & 2,108,797 & 93,803 & 307 & 99 \\
China & 5,979 & 243,840 & 27,431 & 41 & 38 \\
UK & 6,355 & 1,145,060 & 91,260 & 180 & 87 \\

\end{tabular} }
\caption{CIDS country rankings.}
\label{tab:cids}
\end{table}

Table~\ref{tab:scimago} presents the raw numbers of the SCImago country rankings for the studied countries.
Table~\ref{tab:cids} presents the raw numbers obtained by CIDS (see~\ref{app:cids}). Numbers in CIDS are much smaller than 
in SCImago since CIDS analysis only dealt with the top 30 researchers of each country.

\begin{table}[!h]
\centering
\resizebox{0.95\textwidth}{!}{ \begin{tabular}{l|rrrrr}
Country & Citable documents & Citations & Self Citations & Cits per Doc & H index \\
\hline \hline
USA  & 100\% & 100\% & 48\% & 100\% & 100\% \\
China  & 40\% & 9\% & 54\% & 30\% & 28\% \\
UK  & 26\% & 24\% & 24\% & 89\% & 62\% \\
\end{tabular} }
\caption{SCImago country rankings: percentage of USA numbers.}
\label{tab:scimagop}
\end{table}

\begin{table}[!h]
\centering
\resizebox{0.95\textwidth}{!}{ \begin{tabular}{l|rrrrr}
Country & Citable documents & Citations & Self Citations & Cits per Doc & H index \\
\hline \hline
USA  & 100\% & 100\% & 4\% & 100\% & 100\% \\
China & 87\% & 12\% & 11\% & 13\% & 38\% \\
UK & 92\% & 54\% & 8\% & 59\% & 88\% \\
\end{tabular} }
\caption{CIDS country rankings: percentage of USA numbers.}
\label{tab:cidsp}
\end{table}

Table~\ref{tab:scimagop} compares the USA SCImago numbers with the China and UK numbers.
The table shows that China has the second largest number of cited papers and self-citations, and 
UK has the second largest in all the other numbers.

Table~\ref{tab:cidsp} compares the USA CIDS numbers with the China and UK numbers.
The table shows that UK has a larger percentage than China for all the numbers except 
self-citations.

\section{Conclusions}
\label{sec:conclusions}
Comparing CIDS with SCImago results, we see that CIDS was able to give the same big picture as SCImago,
except in the number of cited papers. However, this can be explained by the recent 
growth of China's research leading to a high number of young researchers, and CIDS 
only analyzed the top 30. This also explains why China has not the second largest number of citable documents 
as in SCImago. Even so, China's number of citable documents is very close to UK's number in comparision to the 
their larger difference in the citation numbers.

Even by using a na\"{i}ve selection of team members (using the email suffix),
the results show that CIDS is a feasible alternative to get the big picture of team rankings
using minimal resources. The rankings can be automatically updated since the retrieval and analysis of the 
bibliographic data is fully automated.

%
 \bibliographystyle{plain}
 \bibliography{cids}
%
%
%

\appendix 

\section{CIDS Analysis}
\label{app:cids}
\subsection{USA}
\includegraphics[page=1,width=.9\textwidth]{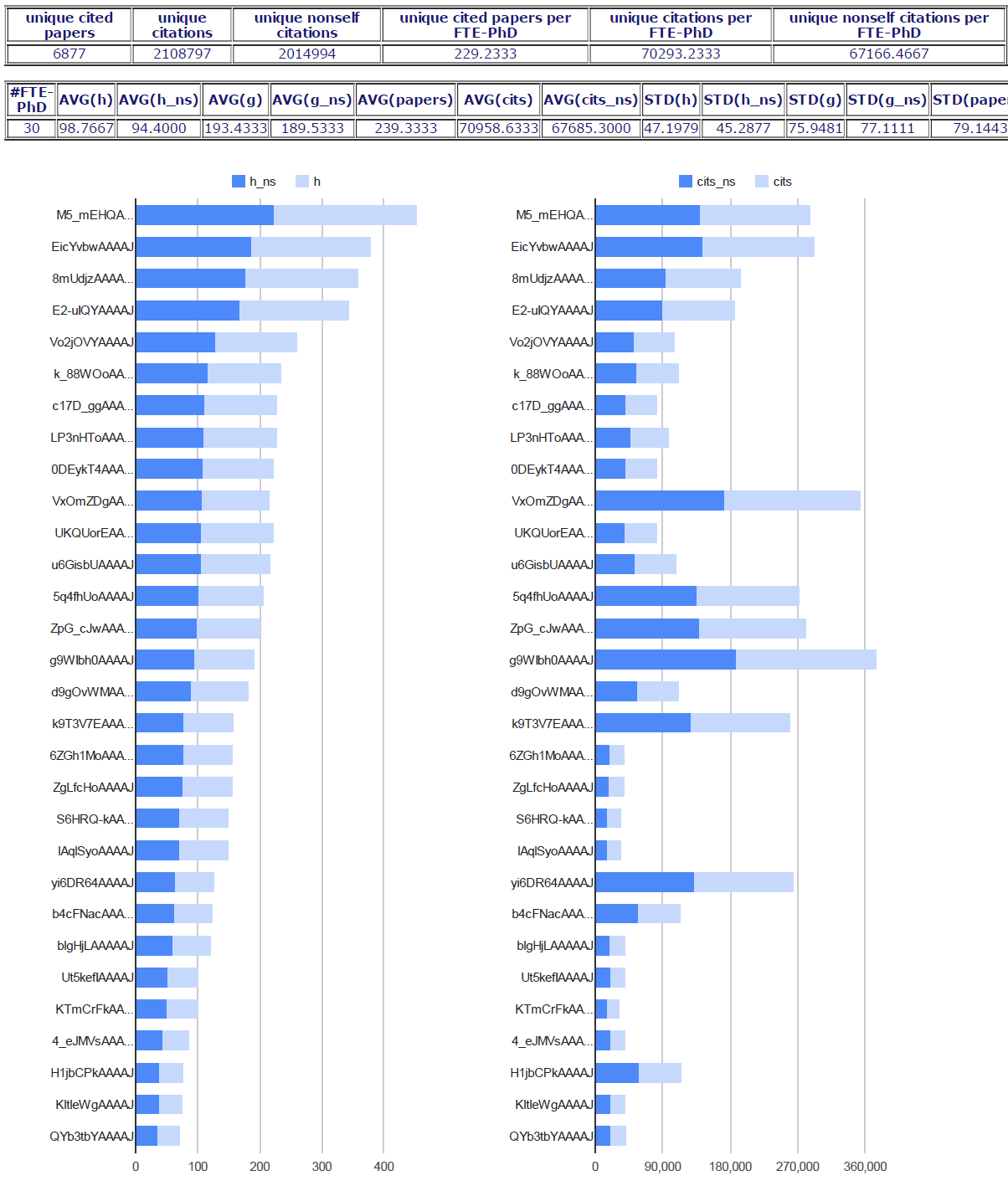}

\subsection{China}
\includegraphics[page=1,width=.9\textwidth]{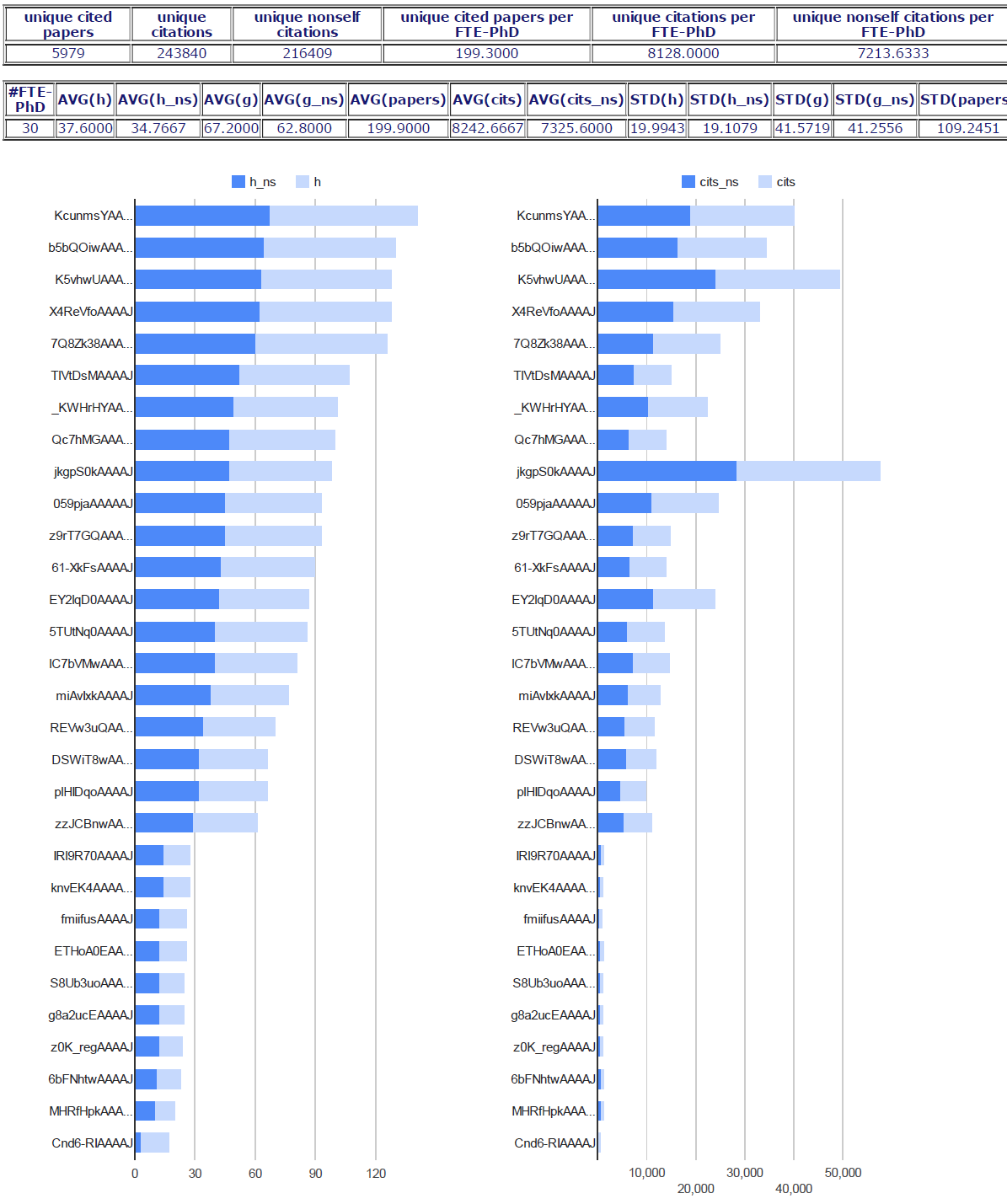}

\subsection{UK}
\includegraphics[page=1,width=.9\textwidth]{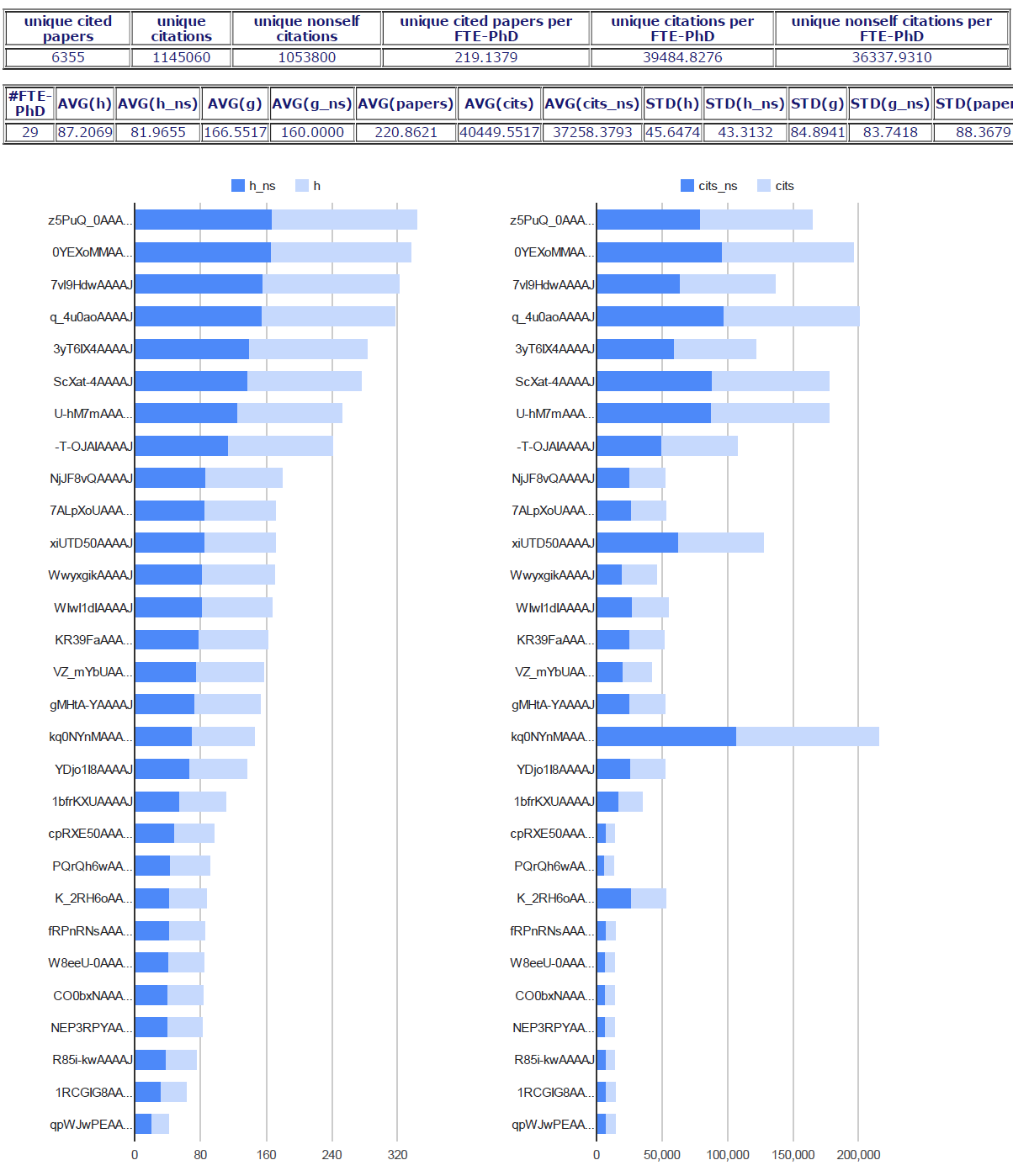}

\end{document}